\documentclass[preprint,authoryear,12pt]{elsarticle}

\usepackage{amssymb}

\journal{Computer Speech and Language}

\begin{document}

\begin{frontmatter}



\title{Glottal Source Processing: from Analysis to Applications}


\author{Thomas Drugman, Paavo Alku, Abeer Alwan, Bayya Yegnanarayana}

\begin{abstract}
The great majority of current voice technology applications relies on acoustic features characterizing the vocal tract response, such as the widely used MFCC of LPC parameters. Nonetheless, the airflow passing through the vocal folds, and called glottal flow, is expected to exhibit a relevant complementarity. Unfortunately, glottal analysis from speech recordings requires specific and more complex processing operations, which explains why it has been generally avoided. This review gives a general overview of techniques which have been designed for glottal source processing. Starting from fundamental analysis tools of pitch tracking, glottal closure instant detection, glottal flow estimation and modelling, this paper then highlights how these solutions can be properly integrated within various voice technology applications.

\end{abstract}


\end{frontmatter}


\section{Introduction}
\label{sec:intro}

Speech is produced as a result of exciting a time-varying vocal tract system by a time-varying excitation. Thus the speech signal carries information about the time-varying speech production system in the relations among the sequence of samples of the signal. This information is captured and interpreted effortlessly by the human auditory perception mechanism, even when the signal is degraded to some extent as in the case of perception of speech in a live room. But the challenge in speech signal processing is to extract the time-varying characteristics of the excitation and the system from the signal by means of computational algorithms. It is essential to extract the production characteristics from the signal, to determine the different types of information embedded in the speech signal, such as speech message, speaker characteristics, emotional state of the speaker, etc. This also reflects the sophistication of the speech production and perception mechanism.
\par
In speech production the major type of excitation is from the glottal source at the larynx, where the airflow from the lungs is intercepted by the normally closed vocal folds, which are held together by the tension on the folds. For sufficiently high pressure from the lungs the tension is not sufficient to hold the folds together, and hence they are forced to open releasing the air through the vocal tract system. But when the pressure is released, the vocal folds tend to close abruptly, causing an impulse-like excitation to the vocal tract system. The opening and closing of the vocal folds take place in a quasi-periodic manner. The nature and frequency of vibration depend on several factors such as the mass and tension on the membranes of the vocal folds, besides the pressure difference on either side of the folds.
\par
The nature of the vocal fold vibration can be controlled voluntarily to produce voices with different qualities such as tense, lax, creaky, breathy, etc. It is this control that enables us to produce speech with different emotion characteristics. The vocal fold vibration also is controlled involuntarily due to the source-system coupling. For example, if the free flow of air through the vocal tract is constrained due to constriction such as in voiced fricatives, the pressure difference across the folds will decrease, thus reducing the frequency of the vibration of the vocal folds. Involuntary control of the vocal folds vibration also takes place when paralinguistic sounds like laughter, cough, sneeze, etc., are produced.
\par
The sophistication of control on glottal vibration is easily felt by the listener, as he/she is able to perceive the subtle changes of the vibrations for each voice quality. But it is extremely difficult to determine the features of vibration that contribute to perception of a given voice quality. It is even more difficult to extract those features from the speech signal.
\par
Study of features of glottal vibration by high speed photography may help to identify the contribution of different parts of the glottis. But it is not possible to collect such data during natural production of sounds, where the vocal tract system also is time-varying. Moreover, visibly significant features of glottal vibration may not constitute significant features in the production of the speech signal, or may not play a significant role in the perception of those sounds. Other measurements such as Electroglottograph (EGG) will only bring out some specific characteristics, such as the impedance across the folds during opening and closing phases in each cycle. Such signals do not give any indication of the pressure changes that eventually produce speech. Of course EGG does give an idea of the open and closed regions of the glottal vibration, and may help as a reference to compare the performance of methods that extract such regions from the speech signal. 
\par
Methods to derive physical or analytical models of glottal vibration help to understand the significance of some features of the glottal vibration \citep{Fant01}. But these methods do not help in extracting actual features of glottal vibration from the speech signal, and hence do not help to identify the features that contribute to the specific voice qualities related to emotion or to some specific speech sounds, such as {\it trills}, for example.
\par
From speech analysis point of view, glottal source processing is examined from modal and non-modal phonation types \citep{Gordon02}. The modal voice represents mostly neutral phonation type with little variation from period to period. It also assumes that there is significant excitation around the glottal closure instant (GCI). Non-modal phonation types involve significant variation in glottal source characteristics. For analysis purposes the different phonation types in normal speech are broadly categorized as shown in Fig. 1.

\begin{figure}[htp]
\centering
\includegraphics[width=14cm, height=3cm]{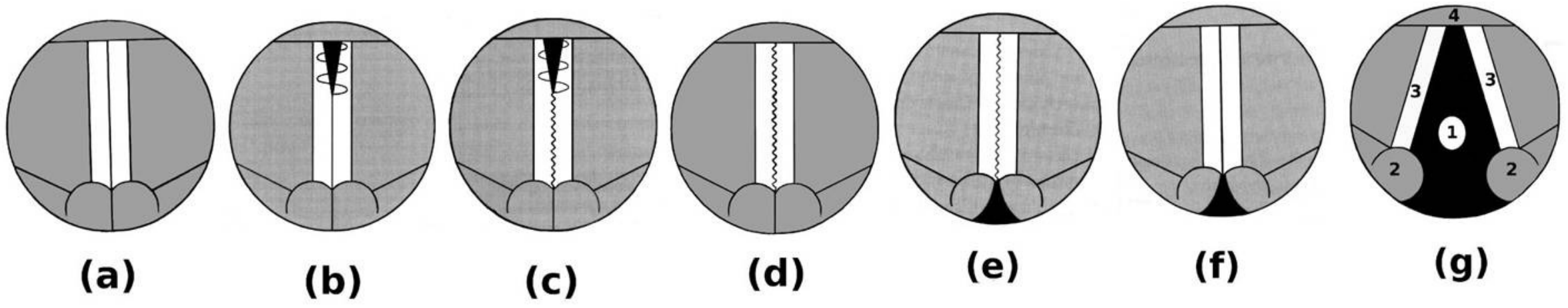}
\caption{\it Stylized glottal configurations for various phonation types: (a) Glottal Stop, (b) Creak, (c) Creaky voice, (d) Modal voice, (e) Breathy voice, (f) Whisper, (g) Voicelessness (1. Glottis, 2. Arytenoid cartilage, 3. Vocal fold and 4. Epiglottis}  
\label{fig}
\end{figure}

\par
Note that only in the case of creaky voice, modal voice and breathy voice, there is vibration of the vocal folds as shown by the corrugated line between the folds. Normally, for glottal source processing, phonation types, where there is vibration of the vocal folds, are considered most of the time. But the figure clearly illustrates that the other four types of phonation indicated in the figure are also important, as they are used in the production of certain types of sound units which have phonemic significance in different languages.

\par
The most important features of the glottal source is the significant impulse-like excitation within each cycle. The nature of the impulse in terms of its sharpness gives an indication of the strength or loudness \citep{Guru06}. In addition, the behavior of the closing phase of glottal vibration, and the presence of other secondary excitations, also contribute to the voice quality of the sound. In order to describe the important features of the glottal excitation, it is necessary to extract not only the GCIs but also the nature of the signal around the GCIs, and the locations and strengths of the secondary excitations. 
\par

Most studies focus on extracting only the GCIs from the speech signal, as it is difficult to derive the other features due to their relatively low amplitudes. Some attempts have been made to determine the secondary excitation instants and their strengths also, attributing them to the instants of glottal opening and to multiple excitations within a cycle \citep{Thomas07}.
\par
 
Other important features of glottal source are the instantaneous fundamental frequency ($F_0$) and the shape of the glottal wave approximated by the parameters representing the main components of glottal wave. Some of the parameters of interest are the opening phase, open phase, closing phase and closed phase regions. The sophistication of different voice qualities is mainly due to variations of these features of the glottal source, which makes glottal source processing an important and challenging study in speech analysis. The objective of this review paper is to present a comprehensive account of various aspects of glottal source processing.
\par
In order to appreciate the sophistication of the glottal source excitation in speech production, Section \ref{sec:physiology} presents an account of the physiology  of glottal production system by describing briefly the organs that contribute to the glottal vibration. Section \ref{sec:synchro} discusses the important features of glottal vibration, namely, the perception of pitch and its variation due to instantaneous fundamental frequency ($F_0$), the speech polarity detection and the instant of significant excitation in each glottal cycle, called the glottal closure instant (GCI). This section also discusses the issues involved in extracting these features form the speech signal, and gives a brief review of the methods to extract these components. Section \ref{sec:source} deals with characterization of glottal wave from the speech signal. The section also considers the parameterization of the extraction glottal source information both in the time-domain as well as in the frequency-domain. The significance of the glottal source processing can be appreciated only when it is studied in the context of different applications. Hence the applicability of the glottal source processing is illustrated in four application scenarios in Section \ref{sec:applicability}. Finally, Section \ref{sec:conclu} gives a summary and conclusions in the form of identifying challenges in glottal source processing from speech collected in practical environments.


\section{Physiology of glottal production}
\label{sec:physiology}

The human speech production system allows a speaker to produce a vast range of sounds. The system consists of many organs intervening in the phonation process, which can be generally categorized into three groups: the lungs, the larynx, and the vocal tract. From a physiological point of view, the airflow from the lungs is pushed through the larynx, where the airflow is modulated by the vibration of the vocal folds (also known as the vocal cords). The vocal folds vibration converts the airflow to acoustic energy and provides an excitation signal to the vocal tract. The vocal tract consists of the oral, nasal, and pharyngeal resonant cavities, which further shapes (or filters) the spectrum of the airflow signal. The resultant airflow is then radiated by the lips. Different aspects of the sound produced can be changed by manipulating the vocal folds vibration pattern, or the configuration of the vocal tract above the larynx (the tongue, jaw, soft palate, and lips).

The ``voice source'' represents the excitation signal produced by the vocal folds. The oscillation of the vocal folds periodically interrupts the airflow from the lungs and creates changes in air pressure \citep{vandenBerg58,Kreiman2011foundations}. When no sound is being phonated, the vocal folds are usually open. To produce unvoiced sounds, the vocal folds are held apart, allowing the airflow to pass through. A noise excitation signal is generated due to the turbulence of the flow. To produce voiced sounds, the muscles which control the closing of the vocal folds (adductor muscles) bring the vocal folds together in order to provide resistance to the air pressure from the lungs. The air pressure below the closed vocal folds (subglottal pressure) forces the vocal folds to open, which allows airflow to pass through the glottis. Then two factors contribute to the closing of the glottis again. The first is the elasticity of the tissue, which forces the vocal folds to regain their original configuration near the midline (the closed position). The second factor is the aerodynamic forces. One such force is described by the Bernoulli Effect, which causes the drop of pressure between the vocal folds when airflow velocity increases. Another aerodynamic force occurs when vortices form in the airflow as it exits the glottis, creating an additional negative pressure between the vocal folds \citep{McGowan88}. Once the vocal folds are closed, the air pressure below them builds up again, and the vocal folds are blown apart to start the process again. This cycle is repeated many times during one second and the cycle duration is called the ``fundamental period'' ($T_0$). Its frequency is referred to as the ``fundamental frequency'' ($F_0$).

According to the linear speech production model \citep{Fant70}, speech signals are generated by filtering the voice source by the vocal tract transfer function. Modeling the glottal source has been an important topic for decades and has applications in many areas, such as speech coding and speech synthesis. Many source models have been proposed with varying levels of complexity, such as the Rosenberg \citep{Rosenberg71}, Liljencrants-Fant (LF) \citep{Fant85}, Fujisaki-Ljungqvist (FL) \citep{Fujisaki86}, and Rosenberg++ (R++) \citep{Veldhuis1998computationally} models. With three parameters, the Rosenberg trigonometric model has two separate functions for the opening and closing phases to represent the glottal flow volume velocity \citep{Rosenberg71}. The LF and FL models represent the first derivative of the glottal volume velocity pulse, which incorporates lip radiation effects. The four-parameter LF model \citep{Fant85} uses a combination of sinusoidal and exponential functions, and is commonly used in speech synthesis. With six parameters and polynomial functions, the FL model provides greater detail in modeling the glottal pulse shape, but the increased number of parameters also makes it more difficult to use in practice. The R++ model \citep{Veldhuis1998computationally} is computationally more efficient but perceptually equivalent when compared to the LF model. Motivated by high-speed images of the vocal folds, a four-parameter glottal flow model, denoted EE1, was introduced \citep{Shue10a}. EE1 uses a combination of sinusoidal and exponential functions similar to the LF model, but with the ability to adjust the slopes of the opening and closing phases separately. Another glottal flow model, denoted EE2, by \citet{Chen2012Model} improves the EE1 model by redefining the model parameters (speed of opening and speed of closing) to allow for lower computational complexity, faster waveform generation, and more accurate pulse shape manipulation. The EE2 model was also used for automatic glottal flow estimation from acoustic speech signals. With the availability of more physiological data, we anticipant further improvements to source modeling.


\section{Tools for Synchronization}
\label{sec:synchro}
Contrarily to the conventional extraction of spectral envelope based features (such as the MFCC or LPC parameters) which is carried out asynchronously, the estimation and parameterisation of the glottal source (as it will be explained in Section \ref{sec:source}) generally requires to process windows whose duration is proportional to the pitch period, as well as the knowledge of the GCI position. This section aims at providing a review of the existing tools necessary for proper synchronization: pitch tracking in Section \ref{ssec:pitch}, speech polarity detection (Section \ref{ssec:polarity}) and finally GCI determination in Section \ref{ssec:GCI}.


\subsection{Pitch Tracking}
\label{ssec:pitch}
The source-filter model of speech production assumes that speech signals can be modeled as an excitation signal filtered by a linear vocal-tract transfer function. The fundamental frequency (F0) is
defined as the inverse of the period of the excitation signal during the voicing state. Accurate F0 tracking in quiet and in noise is important for several speech applications, such as speech coding, analysis and recognition.

Some F0 tracking algorithms are based on the source-filter theory of speech production and estimate F0 for voiced speech segments. They assume that F0 is constant and the vocal tract transfer function is
time invariant within a short period of time, e.g, a frame of 10-20 milliseconds. These algorithms are single-band or  multi-band approaches. There are several methods to generate F0 candidates. SIFT \citep{Abeer-4} applies inverse filtering to voiced speech to obtain the excitation signal from which it estimates F0 by using autocorrelation. Cepstral-based methods (e.g., \cite{Abeer-5}) separate the excitation from the vocal tract information in the cepstral domain by using a homomorphic transformation; the interval to the first dominant peak in the cepstrum is related to the fundamental period. RAPT \citep{Abeer-6} and YAPPT \citep{Abeer-7} generate F0 candidates by extracting local maxima of the normalized cross correlation function which is calculated over voiced speech. Praat \citep{Abeer-8} calculates cross correlation or autocorrelation functions on the speech signal and regards local maxima as F0 hypotheses. TEMPO \citep{Abeer-9} obtains F0 candidates by evaluating the 'fundamentalness' of speech which achieves a maximum value when the AM and FM modulation magnitudes are minimized. YIN \citep{Abeer-10} uses the autocorrelation-based squared difference function and the cumulative mean normalized difference function calculated over voiced speech, with little post-processing, to acquire F0 candidates. \cite{Abeer-11} obtain F0 candidates from exploiting the impulse-like characteristics of excitation in glottal vibrations. Finally, \cite{Abeer-12} simultaneously perform frame-wise F0 candidate generation and time-direction smoothing.

In the multi-band method, a decision module is usually used to reconcile the F0 candidates generated from different bands. \cite{Abeer-13} use measurements of peaks and valleys of voiced speech
as input to six separate functions whose values are then processed by an F0 estimator to obtain F0 candidates. \cite{Abeer-14} calculate autocorrelation functions of the spectral magnitudes in different
bands and then obtain F0 candidates by evaluating the local maxima of the functions. \cite{Abeer-15} detect F0 candidates by minimizing the values of sinusoid-based error functions calculated on 4 frequency
bands. 

Multi-band methods \citep{Abeer-14,Abeer-15} typicaly retain F0 candidates obtained from the most reliable band, while those inspired by Licklider's theory of pitch perception use empirically-based 'soft-decisions' to merge the information from different bands (e.g., \cite{Abeer-18}). In \citep{Abeer-18}, authors showed that integrating AMDF values across different channels in the time domain can improve F0 estimation accuracy. \cite{Abeer-18} used correlograms to select reliable frequency bands, modeled F0 dynamics using a statistical approach, and then searched for the optimal F0 contour in an HMM framework.

Some of the pitch tracking algorithms were designed to be noise robust. For example, a method based on the Summation of the Residual Harmonics (SRH) was proposed in \citep{SRH}. The SRH criterion exploits the harmonic structure of the LP residual excitation both for pitch estimation, as well as for determining the voicing segments of speech. SRH was shown to be particularly robust to additive noise, leading to a significant improvement in adverse conditions over six representative state-of-the-art techniques. Finally, in \citep{Abeer-20}, a Statistical Algorithm for F0 Estimation (SAFE) was proposed utilizing a 'soft-decision' method. A data-driven approach is used to learn how the noise affects the amplitude and location of the peaks in the Signal-to-Noise Ratio (SNR) spectra of clean voiced speech. The likelihoods of F0 candidates were obtained by evaluating the peaks in the SNR spectrum using the corresponding models learned from different bands.


\subsection{Speech Polarity Detection}
\label{ssec:polarity}
When a microphone is used to record a speaker, inverting its electrical connections will cause an inversion of the polarity of the acquired speech signals. The origin of a polarity in the speech signal stems from the asymmetric glottal waveform exciting the vocal tract resonances. During the production of voiced sounds, the glottal source exhibitis periodic discontinuities at GCIs. As described by models of the glottal source (see Section \ref{sec:physiology}), the speech polarity is defined as being positive if the glottal flow derivative exhibits a negative peak at the GCI. Otherwise it is said to be negative.

Our human ear is mostly insensitive to a polarity change \cite{Sakaguchi}. Nonetheless the performance of several speech processing techniques can be severely deteriorated if the signal polarity is erroneous. This is the case for the majority of the GCI detection algorithms which will be described in Section \ref{ssec:GCI}, as well as for the methods of glottal source estimation and parameterization which will be explained in Section \ref{sec:source}. Detecting correctly the speech polarity is then a necessary step to ensure the good behaviour of the aforementioned techniques.

Several approaches have been designed for the automatic detection of the polarity from the speech signal. The idea of the Gradient of the Spurious Glottal Waveforms (GSGW, \cite{GSGW}) technique is to investigate the discontinuity at the GCI in an estimated glottal waveform, whose sign is dependent upon the speech polarity. A criterion is proposed based on a sharp gradient of the spurious glottal waveform near the GCI \cite{GSGW}. In the Phase Cut (PC, \cite{PC}) method, it is searched for the time instant where the two first harmonics are in phase (and which should correspond roughly to the GCI). As their phase slopes are linked by a factor 2, the phase value where they intersect is $\phi_{cut}=\phi_2 - 2 \phi_1$, where $\phi_1$ and $\phi_2$ denote the phase of the first and second harmonics. A value of $\phi_{cut}$ close to 0 (respectively $\pi$) is expected to be due to a positive (respectively negative) peak in the excitation \cite{PC}. The Relative Phase Shift (RPS, \cite{PC}) technique is derived from PC and exploits a greater amount of harmonics. It makes use of Relative Phase Shifts (RPS's), denoted $\theta(k)$ and defined as $\theta(k) = \phi_k - k\cdot \phi_1$, where $\phi_k$ is the instantaneous phase of the $k^{th}$ harmonic. When the excitation exhibits a positive peak, RPS's have a smooth structure across frequency. This smoothness is shown to be dramatically sensitive to a polarity inversion \cite{PC}. For a speech signal, the OMPD algorithm \citep{OMPD} calculates on a sample-by-sample basis statistical moments oscillating at the local fundamental frequency. The key idea of OMPD is to compute two oscillating moments (with an odd and even orders) such that their phase shift allows to determine the correct polarity. Finally, the Residual Excitation Skewness (RESKEW, \citep{RESKEW}) approach exploits the statistical skewness of two excitation signals: the LP residual, and a rough approximation of the glottal source. RESKEW was shown in \citep{RESKEW} to clearly outperform all aforementioned techniques on 10 large speech corpora, and to be more efficient in terms of computational load as well as noise and reverberation robustness.


\subsection{Glottal Closure Instant Detection}
\label{ssec:GCI}

The acoustic pressure variations caused by the glottal vibration to the airflow from the lungs can be viewed as the major excitation component of the speech production system.  This excitation signal is filtered by the response of the vocal tract system to generate the speech signal. Thus the excitation component is convolved with the vocal tract response. If the excitation component is merely impulse-like, then the speech signal corresponds to the response of the vocal tract system. But, as mentioned earlier, the excitation component consists of many features of the glottal vibration, and all of them together contribute to the perception of the voice quality in speech. Among the features of the glottal vibration, the glottal closure instant assumes significance, as the glottal vibration is impulse-like around the GCI, and hence can be considered as the major component of glottal excitation. While methods for extracting the the glottal flow information are desirable to understand the features  of glottal excitation, the currently available methods like zeros of z-transform (ZZT) and complex cepstrum are limited in their use due to choice of the signal region for analysis and also due to their computational cost \citep{Drugman08,Drugman09,Bozkurt10}. Moreover, the assumptions such as mixed-phase model and combination of causal and anti-causal components are not valid for practical speech data, where even slight degradation due to additive noise or reverberant can violate the assumptions. 
\par
Most of the studies in glottal source estimation focused mainly on detection of the GCIs. It is interesting to note that the knowledge of GCIs help in several speech analysis situations such as detection of region of glottal activity \citep{Murty13}, estimation of $F_0$ \citep{Yegna14}, estimation of formant frequencies \citep{Joseph15}, characterization of loudness of speech \citep{Guru06}, analysis of laugh signals \citep{Sudheer04}, and pitch extraction from multi-speaker data \citep{Murty18}. Also GCI-based analysis of speech can be used in several applications such as time delay estimation \citep{Yegna19}, determination of number of speakers from mixed signals \citep{Swamy20}, speech enhancement in single and multichannel cases \citep{Prasanna21}, multi-speaker separation and prosody modification \citep{ksrao23}.

\par
There were many reviews on the glottal flow detection with emphasis on GCI. Paavo summarizes the efforts in GIF methods over different decades, indicating that most of the emphasis was on filtering out the effect of vocal tract resonances to derive the glottal excitation signal \citep{Paavo11}. Wavelet transform has been applied to detect singularities. The GCIs are detected by locating the local maxima of the wavelet transform using a multi-scale product. The concept of lines of maxima amplitude (LoMA) across all the scales in the wavelet transform domain is used for GCI detection \citep{Taun26}. Drugman gives a brief review focused on GCI detection \citep{Drugman28}. In another brief review, the GCI detection over a range of voice qualities was discussed, with emphasis on GCI detection in modal and non-modal phonation \citep{Kane31}. The non-modal phonation display varying glottal source characteristics which include voices like creaky, breathy, tense, harsh and falsetto.

\begin{figure}[htp]
\centering
\includegraphics [width=14cm, height=14cm]{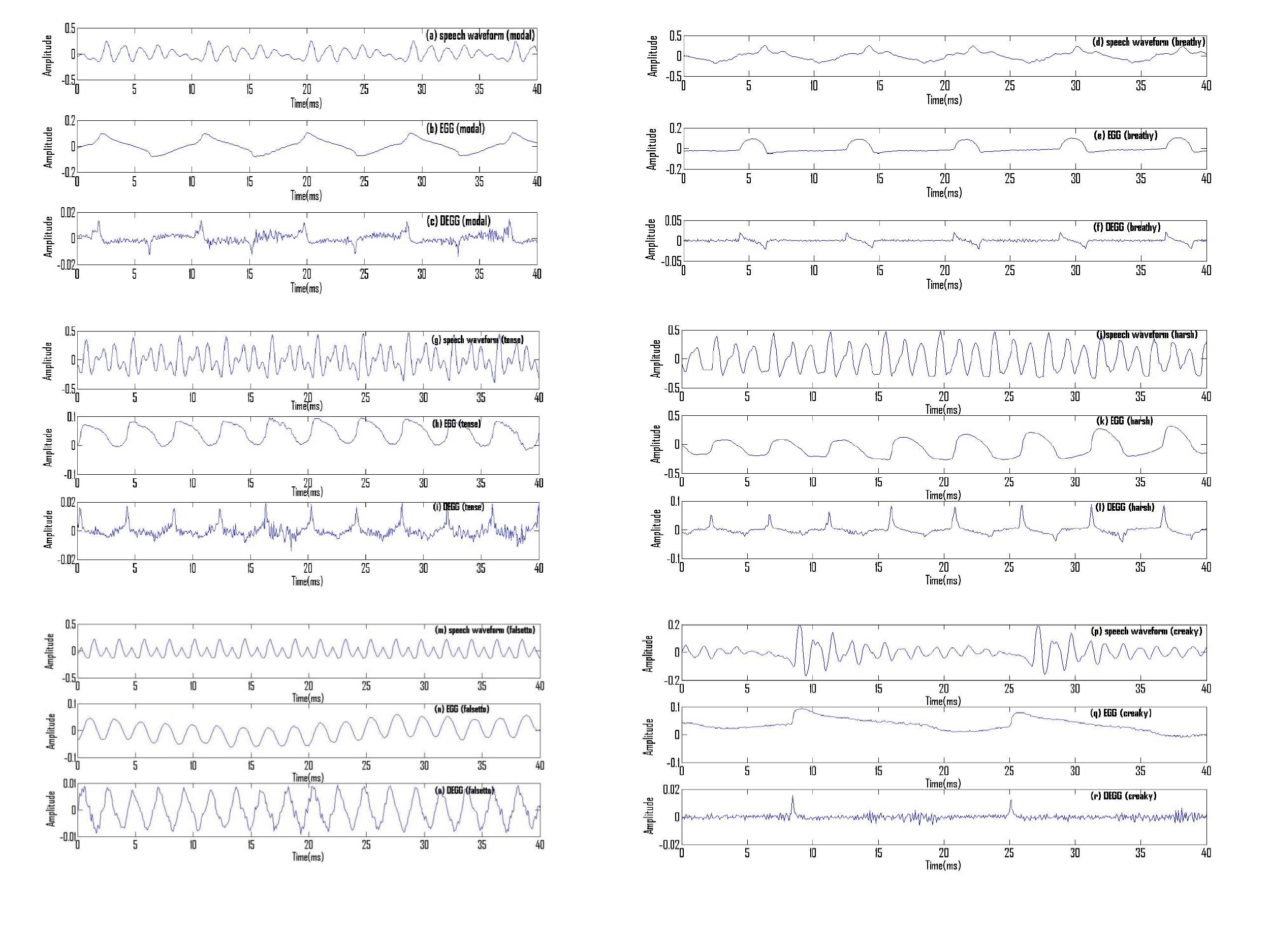}
\caption{\it{Speech waveform, EGG and dEGG for different phonation types}}
\end{figure}

\par
There are several phonation types involving glottal vibration that have distinct characteristics around GCI as shown in Fig. 2. The figure shows the speech waveform, the EGG and the difference EGG (dEGG) for different types of phonation, namely, modal, breathy, tense, harsh, creaky and falsetto and creaky. The purpose of giving these cases is to show the differences in the features of glottal vibration, seen mostly in the dEGG. It is clear that the characteristics of the impulse-like excitation vary in each case, and these characteristics are perceived in speech by a human listener. But it is a challenge to extract such glottal source information from the speech signal.  

\par
In addition, the dEGG for some paralinguistic sounds like laughter and shout show certain distinct characteristics within each period as compared to modal voice \citep{Vinay03}. In particular, for the laughter case one can observe the rapid temporal variations of the glottal excitation features, namely, the instantaneous fundamental frequency ($F_0$) and the strength of excitation \citep{Sudheer04,Sathya05}.

\par
Ideally GCI is a characteristic of the signal produced by the glottal vibration, and hence the signal captured directly like the EGG has the correct information about GCI, its location and shape of the source signal around GCI. Note that EGG does not reflect many of the characteristics of glottal vibration caused by the changes in the acoustic pressure during speech production. But for GCI, the EGG can be used as a reference to evaluate the methods proposed for GCI detection.

\par
As mentioned earlier, it is preferable to have methods which extract GCIs directly from the speech signal. These methods may be grouped into the following five categories:\\
\par

A: Methods which can exploit the property of impulse-like excitation at GCI.\\
B: Methods based on the properties of group-delay.\\
C: Methods based on predictability of all-pole linear predictor.\\
D: Methods based on short-time energy of speech signal.\\
E: Methods based on combination of several techniques.\\

\par
Category A: The methods in this category pass the speech signal through a system which reduces the effects of the response of the vocal tract, at the same time preserve the impulse-like excitation property at GCI. Zero frequency resonator (ZFR) is one such method where the speech signal is passed through a cascade of two ideal digital resonators located at $0$Hz, so that the effects of all higher frequency resonances are reduced significantly \citep{Murty32}. The operation of ZFR is like integrating the signal four times, thus producing a smoothed signal which preserves the impulse excitation characteristics in the small fluctuations of the ZFR output. The trend in the ZFR output is removed by subtracting the average over a window of the size in the range of 1 to 2 pitch periods. The resulting  mean subtracted signal, called zero frequency filtered (ZFF) signal gives the GCI locations precisely at the instants of negative to positive zero crossings. The ZFR approach gives the GCIs even for aperiodic sequence of impulse-like excitations. The ZFR approach also gives GCIs even when the signal is degraded with additive or babble noise. However, if there are degradations due to channel effects, the speech signal needs to be transformed into Hilbert envelope (HE) of LP residual before applying the ZFR method to capture the impulse-like excitation characteristics.
\par
Category B: Group-delay is the negative derivative of the phase of the Fourier transform of a signal. For this, a segment of windowed signal need to be considered. The average of the group-delay function for each segment corresponds to the delay of the impulse in the segment from the center of the segment. Thus the average of the group-delay function computed at every sampling instant has a zero value at the GCI, i.e., when the GCI is located at the center of the analysis segment \citep{Smits33}. Thus, by this method, the point property of the GCI is obtained by using the block processing approach needed for  computation of the group-delay function. Several refinements on this approach lead to algorithms like DYPSA, which uses the zero crossings of the phase slope function derived form the energy weighted group-delay to obtain the candidates for GCI, and these candidate GCIs are refined by employing a dynamic programming algorithm \citep{Naylor34}.
\par
Category C: Many methods for GCI detection rely on the discontinuities in the output of linear prediction model of speech production \citep{Strube35}. An early approach used as a predictability measure to detect the GCI by finding the maximum of the determinant of the autocovariance matrix of the speech signal. Most methods assume that predictability is least around GCI, as reflected by the large amplitude fluctuations in the LP residual. 
\par
The GCI detection from the LP residual is improved by computing its HE, where unambiguous peaks are obtained around GCIs for clean signal, and these peaks match with the sharp discontinuities seen in the differenced EGG \citep{Ananthapadmanabha36}. There are several refinements suggested to improve the GCI detection from the LP residual. A recent one is based on filtering the LP residual with a resonator located at approximately at $F_0$ \citep{Kane31}.
\par
One of the difficulties in using the prediction error for GCI detection is that the LP residual often contains effects due to resonances of the vocal tract system, as the inverse filter does not cancel out the resonances completely. Moreover, the inverse filter basically emphasizes the low signal to noise ratio (SNR) regions of the high frequency, and thus reducing the SNR in the LP residual. This reduction in SNR results in the loss of resolution of the impulse-like behavior around the GCI.
\par
Category D: Some of the early methods for GCI detection are based on short time energy of the speech signal or from the features in its time-frequency representation \citep{Jankowski38,Ma39}. Note that the time-frequency representation or energy computation  require block processing, and hence the GCIs cannot be detected accurately. These methods are grouped under the category D. 
\par
Category E: Currently methods are being explored which combine several different methods used for GCI detection. The Yet Another GCI Algorithm (YAGA) is one such method which uses iterative adaptive inverse filtering, wavelet analysis, the group-delay function and M-best dynamic programming \citep{Thomas30}. The method was also used for estimating the glottal opening instants.
\par
SEDREAMS algorithm is another method which uses the mean-based signal to determine the short intervals where GCIs are expected to occur \citep{Drugman28}. The mean-based signal is derived from speech using a window of the size about 1.75 times the mean pitch period, and the resulting signal oscillates at the local pitch period. Using these short intervals at each pitch period, the GCI locations are refined in the LP residual signal. The SEDREAMS algorithm is further refined to handle GCIs from signals produced from different phonation types. The refinement involves applying a dynamic programming method to select the optimal path on GCI locations based on both the strength of the LP residual peak and a transition (from one GCI location to the next) cost. A further post-processing is used to reduce the false peaks. The post-processing uses the output of the LP residual through a resonator located at a frequency corresponding to mean $F_0$.
\par
This brief review shows that there is continuing effort on finding methods for GCI detection from speech signal which are not only accurate (with reference to the ground truth provided by dEGG), but also work for speech signals produced with different phonation types and for paralinguistic signals like laugh, cough, etc. The biggest challenge is to develop methods for GCI detection for speech signals collected in a practical environment, such as in a live room with additive noise and small reverberation. A further challenge is to detect the other minor excitations within a glottal cycle, as all the excitations (major and minor) together contribute to the perception of voice quality in speech. Until these glottal features reflecting different voice qualities are understood and extracted, it is hard for the speech researchers to understand the components of the glottal source that contributes to the expressive or emotional speech.



\section{Glottal Source Estimation and Parameterization}
\label{sec:source}

Relying on the fundamental tools of synchronization which have been presented in Section \ref{sec:synchro}, several methods have been proposed for the automatic estimation of the glottal flow from audio recordings, as well as for its parameterization. These techniques are now reviewed in this section.

\subsection{Glottal Source Estimation}
\label{ssec:estimation}

The article by (\cite{Miller59}) is regarded as the first publication in which Glottal Inverse Filtering (GIF) is used as a method to estimate the glottal source. This study was followed by several publications by Fant and his colleagues (\cite{Fant61}, \cite{Fant62}), as well as by, for example, (\cite{Mathews61}). In these early days, the inverse model of the vocal tract consisted of lumped analog elements which were adjusted based on visual observations of the filter output via an oscilloscope.  Already in these early GIF studies the tuning of the anti-resonances was done by searching for such settings that yielded maximally flat closed phase of the glottal pulseform (\cite{Lindqvist64}). This criterion has since been used in several GIF studies in determining optimal inverse filter settings.

To the best of our knowledge, the first study utilizing digital signal processing (DSP) in the estimation of the glottal source was published by (\cite{Oppenheim68}). Their work actually addressed a more general tool of DSP, the so called homomorphic analysis in which the convolved signal components are transformed into additive components using the cepstral analysis. Separation of speech into the glottal source and vocal tract was studied as an example of the cepstral analysis, hence presenting the first DSP-based experimental results of GIF. Digital cepstral analysis has been later used by, for instance, (\cite{Drugman09}) as the means to separate the glottal source from the vocal tract. Another early study using digital GIF analysis was published by (\cite{Nakatsui70}). Their investigation is the first study which takes advantage of digital filtering as a means to implement the anti-resonances of the vocal tract inverse model.

\cite{Rothenberg73} studied a GIF technique that is based on inverse filtering the volume velocity waveform recorded in the oral cavity rather than the acoustic speech pressure signal captured in the free field outside the mouth. He introduced a special pneumotachograph mask which is a transducer capable of measuring the volume velocity at the mouth. The recorded signal was then subject to inverse filtering, where analog antiresonances were determined using a spectrographic analysis. The use of the pneumotachograph mask benefits from providing an estimate for the absolute DC level of the glottal air flow, an issue that cannot be achieved with GIF techniques that use the free field microphone recording as the input. Even though the original work reported by (\cite{Rothenberg73}) involved the use of analog filters in cancelling the vocal tract resonances, the use of the pneumotachograph mask has later been combined with digital inverse filtering (e.g. \cite{Granqvist03}).

One of the most widely used GIF methods, the closed phase (CP) covariance analysis was proposed by (\cite{Strube74}). The method was further developed a few years later by (\cite{Wong79}) in their article that is one of the most cited papers in the study area. The CP analysis uses linear prediction (LP) with the covariance criterion as a tool to compute digital vocal tract models. By positioning the analysis window of LP into the glottal closed phase (i.e. the time span during which there is no contribution from the excitation), the CP analysis is capable of separating the glottal source and the vocal tract. In comparison to old analog techniques, the use of LP introduced a notable improvement in GIF because the vocal tract model adjusts automatically to the underlying speech signal. The CP analysis has been successfully used in many voice production studies (e.g. \cite{Veeneman85}, \cite{Krishnamurthy86}). Other investigations, however, have demonstrated that the method is sensitive to the extraction of the closed phase position, and even small errors might result in severe distortion of the estimated glottal excitation (e.g. \cite{Riegelsberger93}, \cite{Yegna98}). This drawback can be partly alleviated by using the electroglottography (EGG) instead of the acoustic speech signal in extracting glottal closure instants (e.g. \cite{Krishnamurthy86}). In addition, the performance of CP has been reported to improve by using techniques such as the involvement of speech samples from consecutive cycles (e.g. \cite{Plumpe}, \cite{Yegna98}), the use of adaptive high-pass filtering (\cite{Akande05}), and the use of constrained LP in modelling of the vocal tract (\cite{Alku09}). 

The idea of computing a parametric all-pole model of the vocal tract jointly with a source model was proposed by (\cite{Milenkovic86}) as a basis of his GIF technique. This approach enables utilizing speech samples over the entire fundamental period in the optimization of the vocal tract, an issue which is not fulfilled in the basic form of the CP analysis. The joint optimization of the vocal tract and glottal source has since been used by several authors. Some of these works have utilized the autoregressive model with an exogenous (ARX) input, where the input has been represented in a pre-defined parametric form by using, for example, the Rosenberg-Klatt model or the Liljencrants-Fant (LF) model of the voice source (e.g. \cite{Isaksson89}, \cite{Kasuya99}, \cite{Frohlich01}, \cite{Fu06}, \cite{Berezina10}, \cite{Ghosh11}).

A straightforward and automatic GIF method, named Iterative Adaptive Inverse Filtering (IAIF), was proposed by one of the present author in (\cite{Alku92}). This method estimates the contribution of the glottal excitation on the speech spectrum with a low order LP model that is computed with a two-stage procedure. The vocal tract is then estimated using either conventional LP or discrete all-pole modelling (DAP), an all-pole modelling method that utilizes the Itakura-Saito distortion measure (\cite{El-Jaroudi91}). In the GIF method proposed by Bozkurt and his co-authors, however, a different approach was used (\cite{Bozkurt05}, \cite{Bozkurt07}, \cite{Sturmel07}). Their method, Zeros of Z-Transform (ZZT), does not utilize the LP analysis in the estimation of the source-tract separation, but rather expresses the speech sound with the help of the z-transform as a large polynomial. The roots of the polynomial are separated into two patterns, corresponding to the glottal excitation and the vocal tract, based on their location from the unit circle. A similar kind of separation based on causal-anticausal decomposition has also been used by (\cite{Drugman11}, \cite{Drugman12}) by using the complex cepstrum. Separation based on phase characteristics (i.e. minimum-phase for the vocal tract and mixed-phase for the glottal source) was studied by (\cite{Degottex11}). Their method fits the phase spectrum of the LF model to that of the observed signal at harmonic components by minimizing the mean square phase difference. The technique proposed is different from the true GIF techniques (e.g. CP, IAIF) in the sense that the glottal source signal is not explicitly computed but model parameters are rather solved.

\subsection{Glottal Source Parameterization}
\label{ssec:parameterization}
Analysis of speech production with GIF consists typically of two stages. In the first one, glottal flow waveforms are estimated using GIF by adopting, for example, some of the techniques presented in the previous section. In the second stage, the estimated waveforms are parameterized by expressing their most important properties in a compressed numerical form. In the following, some of the main parameterization techniques developed are shortly described by first discussing the time-domain techniques and then the frequency-domain methods. 

Parameterization of the time-domain glottal flow signals can be computed by using time-based or amplitude-based methods. In the former, the most straightforward classical method is to compute time-duration ratios between different phases (i.e. opening, closing and closed phase) of the glottal flow pulse (see Fig. 1) (e.g. \cite{Timcke58}, \cite{Monsen77}, \cite{Sundberg99}).

\begin{figure}[htp]
  \begin{center}
   \includegraphics[width=12cm]{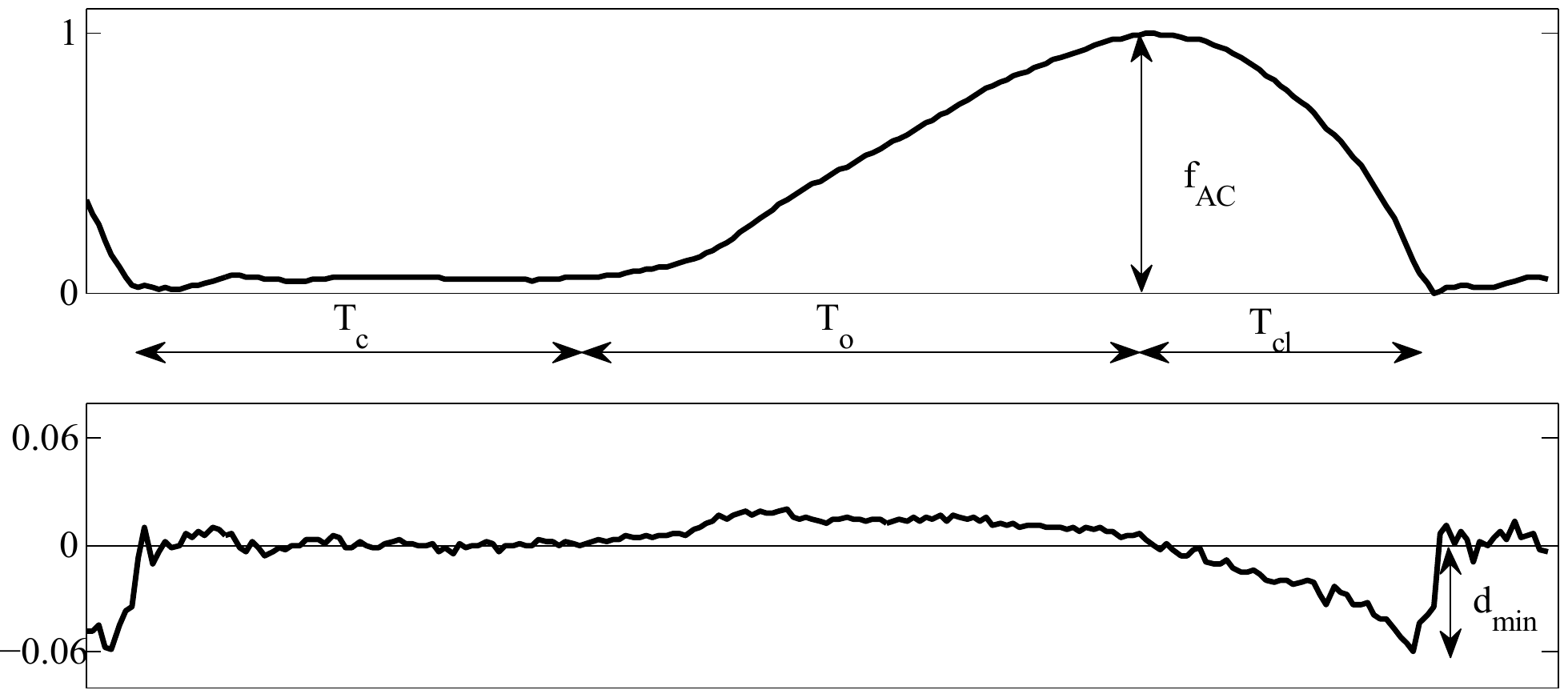}
  \end{center}
\vspace{0cm}
\caption{Time-based parameterization of the glottal source from the flow pulse (upper) and its derivative (lower panel). The flow pulse is divided into three parts: the closed phase ($T_c$), the opening phase ($T_o$), and the closing phase ($T_{cl}$). Most widely used time-based parameters are defined as follows: open quotient ($OQ = \frac{T_o+T_{cl}}{T})$), speed quotient ($SQ = \frac{T_o}{T_{cl}}$), and closing quotient ($ClQ = \frac{T_{cl}}{T}$), where the length of the fundamental period is denoted by $T = T_c+T_o+T_{cl}$. Time-based parameterization can also be computed from amplitude-domain values with $NAQ = \frac{f_{AC}}{d_{min}\cdot T}$).}
\label{fig:Alku}
\vspace{0cm}
\end{figure}

Time-based measures described above are vulnerable to distortions that are present in glottal flow waveforms due to incomplete cancelling of formants by the inverse filter. Therefore, computation of the time-based parameters is sometimes performed by replacing the true opening and closure instants by the time instants when the glottal flow crosses a level which is set to a given value between the minimum and maximum amplitude of the glottal cycle (e.g. \cite{Dromey92}). In addition, time-based features of the glottal source can be quantified by measuring the amplitude-domain values of the glottal flow and its derivative (\cite{Fant95}, \cite{Alku96}, \cite{Alku02}). One such measure, the Normalized Amplitude Quotient (NAQ), was proposed by (\cite{Alku02}). Computation of NAQ is done from two amplitude-domain values, the AC-amplitude of the flow and the negative peak amplitude of the glottal flow derivative (see Figure \ref{fig:Alku}). Since these two amplitude measures are the largest values of the flow and its derivative in a glottal cycle they are straightforward to be extracted even though the estimated glottal sources were distorted. In addition to NAQ, other classical time-based measures can also by represented similarly as shown by (\cite{Gobl03}).

When GIF is computed with the flow mask as proposed by (\cite{Rothenberg73}), it is possible to parameterize the amplitude-domain properties of the time-domain waveforms of the glottal flow and its derivative.  The most widely used amplitude parameters are the minimum flow (also called the DC-offset), the AC-flow and the negative peak amplitude of the flow derivative (also called the maximum airflow declination rate). In addition, the ratio between the AC and DC information has been used in the amplitude-based quantification of the glottal source (\cite{Isshiki81}). 

It is also possible to parameterize the time-domain voice source by searching for an artificial waveform that matches the computed glottal excitation (or its derivative) (e.g. \cite{Strik93}, \cite{Li12}). There are many such artificial glottal source models (\cite{Fujisaki86}) that have been developed during the past three decades ranging from the two-parameter polynomial model proposed by Klatt (\cite{Klatt87}) to more complicated models such as the Liljencrants-Fant model (LF-model) (\cite{Fant85}). Time-domain reproduction of the glottal excitation estimated by inverse filtering is also possible by searching for the physical parameters (e.g. vocal fold mass and stiffness) of the underlying oscillator rather than explaining the waveform of the volume velocity signal (\cite{Drioli05}, \cite{Avanzini08}). 

Instead of fitting the computed glottal flow with a pre-defined function, a parameterization scheme based on a data-driven approach was recently proposed by (\cite{Gudnason12}). More specifically, a glottal flow estimate was first computed with IAIF and the obtained waveform was then processed with Principal Component Analysis (PCA) in order to achieve dimensionality reduction. A set of the first PCA components were then modelled with Gaussian Mixture Models (GMMs). The obtained GMMs were shown by (\cite{Gudnason12}) to enable parameterizing source features, such as non-flatness of the closed phase, that traditional approaches fail to model. Finally, a new parameterization method of the voice source was recently proposed by (\cite{Kane13}) by adopting dynamic programming using which the settings originating from a manual voice source analysis could be combined into an automatic, machine-based analysis.

Frequency domain parameterization of the glottal source has been computed by measuring the spectral skewness with the so-called alpha ratio, which is the ratio between spectral energies below and above a certain frequency limit (typically 1.0 kHz) (\cite{Frokjaer73}). In addition, several studies have quantified the spectral decay of the glottal source by utilizing the level of the fundamental frequency (F0) and its multiple integers, the harmonics. One such measure, called Harmonic Richness Factor (HRF), was developed by (\cite{Childers91}). HRF is defined from the spectrum of the estimated glottal flow as the ratio between the sum of the amplitudes of harmonics above the fundamental and the amplitude of the fundamental. In addition, levels of spectral harmonics have been used in parameterization of the glottal flow by (\cite{Howell92}), who measured the decay of the voice source spectrum by computing linear regression analysis over the first eight harmonics. \cite{Titze92} analyzed the spectral decay of the voice source of singers by computing the difference, denoted by H1-H2, between the amplitude of the fundamental and the second harmonic. A similar measure of the spectral tilt is sometimes computed directly from the radiated speech signal by correcting the effects of vocal tract filtering without computing the glottal flow by GIF (e.g. \cite{Iseli07}, \cite{Kreiman12}). \cite{Alku97} proposed a frequency domain method based on fitting a low-order polynomial into the pitch synchronous source spectrum. Apart from measuring the spectral decay, it is also possible to quantify the glottal excitation by measuring the ratio between the harmonic and non-harmonic components (e.g. \cite{Murphy99}). This approach is justified especially in the analysis if disordered voices, in which the glottal excitation typically involves an increasing amount of aperiodicities due to jitter, shimmer, aspiration noise, and changing of the pulse waveform.


\section{Applicability of Glottal Source Processing}
\label{sec:applicability}

The analysis techniques developed in Sections \ref{sec:synchro} and \ref{sec:source} allow the extraction of glottal source-related features. We will now see how this information can be relevantly incorporated within various voice technology applications: parametric speech synthesis in Section \ref{ssec:synthesis}, speaker recognition in Section \ref{ssec:reco}, biomedical applications in Section \ref{ssec:biomedical} and finally expressive speech analysis and synthesis in Section \ref{ssec:expressive}. The integration of glottal information within these applications is expected to be helpful as it should be complementary with the vocal tract response.

\subsection{Speech Synthesis}
\label{ssec:synthesis}

The great majority of existing techniques for parametric speech synthesis rely on a source-filter model. In this framework, two options are possible according to what is considered to be source and the filter. In the first case, the source is the glottal excitation as physiologically produced by the vocal folds, and the filter refers to the vocal tract response. Beyond the physiological motivation, this approach has the advantage to be more flexible as proper modifications of the glottal contribution are expected to reflect changes of voice quality. Nonetheless, the main drawback of this option is the requirement to reliably and accurately estimate and model the glottal component. In the second case, the filter corresponds to the spectral envelope of the speech signal and the excitation source is the residual signal obtained by inverse filtering after removing the spectral envelope contribution. The residual signal has the advantage to be easily obtained, however its amplitude spectrum is by definition flat and the information about the glottal spectral shaping is lost in the filter component. Therefore its flexibility for speech modifications is much more limited. 

Based on these two approaches, several methods have been proposed to improve the naturalness in HMM-based speech synthesis (\cite{HTS}). Indeed the basic vocoder used in HMM-based synthesis assumes the excitation signal to be either a pulse train in voiced segments, and a white noise in unvoiced regions. This too simple reprensentation causes a typical \emph{buzziness} in the generated speech, as it was found in the old LPC-based speech coders. To overcome this issue, a more elaborated source modeling is required. In (\cite{Yoshimura}), a Mixed Excitation (ME) is proposed to model the residual signal. The ME is the sum of both periodic and aperiodic components which are controlled by bandpass voicing strengths. These latter parameters are trained by HMMs and generated at synthesis time. In a similar way, a ME consisting of a set of high-order state-dependent filters derived through a closed-loop procedure was proposed in (\cite{Maia}). In \cite{Drugman-ICASSP09}, an hybrid approach makes use of a codebook of pitch-synchronous residual frames which are selected at synthesis time, comparably to what is achieved in the Code Excited Linear Prediction (CELP, \cite{CELP}) method. In (\cite{DSM_IS}) and (\cite{DSM_TASLP}), the same authors propose the Deterministic plus Stochastic Model (DSM) of the residual signal. The DSM consists of two contributions acting in two distinct spectral bands delimited by a maximum voiced frequency. Both components are extracted from an analysis performed on a speaker-dependent dataset of pitch-synchronous residual frames. The deterministic part models the low-frequency contents and arises from an orthonormal decomposition of these frames. As for the stochastic component, it is a high-frequency noise modulated both in time and frequency. All these techniques modeling the residual signal have been shown, compared to the traditional pulse excitation, to provide a significantly higher naturalness after HMM-based speech synthesis.

In parallel, several attempts have been carried out to integrate a glottal source modeling in HMM-based speech synthesis. The approach described in (\cite{Cabral07}) incorporates the LF model so as to reduce the \emph{buzziness} and enhance the flexibility. A similar approach was proposed in (\cite{Lanchantin}) where a new glottal source and vocal-tract separation method is used. Finally, a library of natural  glottal waveforms obtained using IAIF is presented in (\cite{Raitio}). Selected glottal pulses are then further interporlated and concatenated. Again, these latter methods were shown to outperform the traditional excitation in the frame of HMM-based speech synthesis.

Besides the application of statistical parametric synthesis, several systems have targeted voice transformation by processing the excitation signal. In this way, (\cite{Cabral08}), (\cite{Degottex_breathy}) and (\cite{ARX}) proposed to use the LF model to perform voice modifications (e.g. in terms of the breathiness or tenseness). Several approaches have also focused on the manipulation of the excitation signal to carry out high-quality pitch modification (\cite{Cabral05}, \cite{Degottex_breathy}, \cite{Drugman_Pitch}). Finally, the glottal source has also been employed in the context of voice conversion (\cite{Childers_VC}, \cite{DelPozo_VC}) where, in addition to improving the main quality, it also offers the possiblity to apply voice quality modifications.

\subsection{Speaker Recognition}
\label{ssec:reco}
The baseline approach used in speaker recognition systems relies on the standard MFCC features. This way of doing consequently discards the possible speaker-dependent information contained in the excitation source. However, as it will be shown in this section, the use of this latter signal could be beneficial as speakers have a different larynx and use their vocal folds in a different way. Furthermore, the speaker-dependent information contained in the glottal flow is expected to be complementary with the vocal tract information.

Here again, several methods have been proposed in the literature and differ depending on whether they use an estimate of the glottal source or the residual signal obtained by inverse filtering after removing the contribution of the spectral envelope (generally achieved by a standard LP analysis).

Several studies have underlined the fact that the residual signal conveys an interesting amount of information regarding the speaker identity. In (\cite{Yegna_AANN}), the residual features are captured
implicitly by a feedforward autoassociative neural network (AANN). The study demonstrates the complementary nature of the residual signal with the standard MFCCs. This approach was further developed in (\cite{Prasanna_AANN}) where, even though the residual information provided lower recognition rates, it was shown to require less amount of training and testing data and achieve better speaker identification when complemented with MFCCs. The work presented in (\cite{Murty_Phase}), which exploits the residual phase, further emphasized the evidence indicating that speaker-specific excitation information is present in the residual signal and that it is complementary with MFCCs. In (\cite{Pati}), the feature extraction from the residual signal is carried out by a non-parametric vector quantization. This method is again shown to be of interest when combined with conventional MFCCs. The same conclusion was drawn in (\cite{Chetouani}) where temporal and spectral features of the residual signal were proposed. Temporal characteritics are based on an auto-regressive modeling making use of second and third-order statistics to account for the non-linear nature of speech signals. Regarding the frequential approach, authors exploit a filter bank method measuring the spectral flatness over the sub-bands. Finally, the approach described in (\cite{Drugman_Signatures}) and (\cite{DSM_TASLP}) propose so-called \emph{glottal signatures} derived from the Deterministic plus Stochastic Model (DSM) of the residual signal. Each speaker is characterized by two signatures corresponding to both deterministic and stochastic components. This technique is shown to clearly outperform other speaker identification approaches based on glottal features.

Glottal flow estimates have also been proven to be useful for speaker recognition tasks. The first attempts were made in (\cite{Plumpe}). For this, authors propose to estimate the glottal flow derivative using a variant of the CPIF method. Glottal source estimates are modeled using the LF model to capture its coarse structure, while the fine structure of the flow derivative is represented through energy and perturbation measures. In (\cite{Slyh}), the Fujisaki and Ljungqvist model was used in conjunction with closed-phase analysis to yield glottal features for a speaker recognition system. In (\cite{Gudnason}), a proper estimate of the vocal tract response is achieved by closed phase LP analysis. The glottal features then consist of the subtraction of the speech cepstral coefficients with those of the estimated vocal tract function. Finally, a similar approach relying on a cepstral representation of the voice source is proposed in (\cite{Kinnunen}) and exploits an IAIF-based separation of the vocal tract and glottal components. Across all these studies, the use glottal features alone did not provide better performance compared to the standard MFCCs, but their combination always led to a reduction of speaker identification errors.

\subsection{Biomedical Application}
\label{ssec:biomedical}
At the crossroad between biomedical engineering and signal processing, the glottal information has been demonstrated to be useful in several healthcare applications. Its most straightforward applicability is probably for voice disorder detection. Indeed speech pathologies are often associated with a dysfunction at the vocal folds level: polyp, oedema, nodule, etc. Features describing the glottal behaviour are therefore of great interest for the automatic audio-based detection of voice pathologies. Jitter and shimmer are two usual measures employed to characterize irregularities in a quasi-periodic signal (\cite{Lieberman}). The jitter refers to the change in the duration of consecutive glottal cycles, while the shimmer reflects the changes in amplitude of consecutive glottal cycles. The study described in (\cite{Silva}) focused on a comparative evaluation of different methods to estimate the amount of jitter in speech signals with a target on their ability to detect pathological voices. Results showed that there were significant differences in the performance of the various algorithms, the best techniques being not necessarily the most commonly used. In (\cite{Vasilakis}), the use of the spectral short-term jitter estimator is proposed to discriminate voice pathologies in running or continuous speech, leading to an interesting performance. Jitter values were found to confirm studies showing a decrease of jitter with increasing fundamental frequencies, and the more frequent presence of high jitter values in the case of pathological voices as time increases. A variant of the traditional HNR measure, called glottal-related HNR (GHNR') is proposed in (\cite{MurtyPatho}) to overcome the F0 dependency in the usual HNR formulation. Within the frame of voice pathology discrimination, GHNR' is shown to provide statistically significant differentiating power over a conventional HNR estimator. The study described in (\cite{Drugman-Patho1}) investigates the use of the glottal source estimation as a means to detect voice disorders. Three sets of features are proposed, depending on whether they are related to the speech or the glottal signal, or to prosody. Results indicate that speech and glottal-based features are relatively complementary, while they present some synergy with prosodic characteristics. Combining glottal with speech-based features provides an interesting discrimination ability. In (\cite{Gomez}), a biometric signature based on the speaker's power spectral density of the glottal source and related to the vocal fold cover biomechanics is presented. The detection capability of the methodology is illustrated on a study case with a control subset of 24 + 24 subjects is used to determine a subject's voice condition in a pre- and post-surgical evaluation. Automatic pathology detection is mentioned as a possible application, but no specific reults are given on this topic. Finally, several phase-based features have been proposed in (\cite{Drugman-Patho2}) to detect voice disorders. The phase information is mostly linked with the glottal production, and it was shown to be particularly well suited to highlight irregularities of phonation compared to its magnitude counterpart.

Glottal characterization has also been shown to be helpful in another biomedical problem: the classification of clinical depression in speech. In (\cite{Ozdas}), the vocal jitter and glottal flow spectrum are proposed as possible cues for assessment of a patient's risk of committing suicide. Three groups of 10 subjects each were considered: high-risk near-term suicidal patients, major depressed patients, and nondepressed control subjects. The mean vocal jitter was found to be a significant discriminator only between suicidal and nondepressed control groups, while the slope of the glottal flow spectrum was a significant discriminator between all three groups. The approach explained in (\cite{MooreDepressed}) also tackles the issue of discriminating depressed speech. It is shown that the combination of glottal and prosodic features produced better discrimination overall than the combination of prosodic and vocal tract features. It was also underlined that glottal descriptors are vital components of vocal affect analysis. 

Finally, the use of the glottal source can be of interest for laryngectomees. Indeed, patients having undergone total laryngectomy cannot produce speech sounds in a conventional manner because their vocal folds have been removed. Gaining a new voice is then the major goal of the post surgery process. Several approaches have targeted the resynthesis of an enhanced version of alaryngeal speech, in order to improve its quality and intelligibility. For this purpose, it is required to re-create an artifical excitation signal based on our understanding of the glottal production. This is generally achieved by using the LF model, as in the two following studies. In (\cite{Qi}), authors resynthesized female alaryngeal words with a synthetic glottal waveform and with smoothed and raised F0. It was shown that the replacement of the glottal waveform and F0 smoothing alone produced most significant enhancement, while increasing the average F0 led to less dramatic improvement. The speech repair system proposed in (\cite{Arantza}) resynthesizes alaryngeal speech using a synthetic glottal waveform, reduces its jitter and shimmer and applies a spectral smoothing and tilt correction algorithm. A subjective assessment reveals a reduction of the perceived breathiness and harshness of the voice. Finally, the solution described in (\cite{McLoughlin}) interestingly focuses on the speech reconstruction from whispered voice, and proposes a modified version of the CELP vocoder.

\subsection{Expressive Speech Analysis and Synthesis}
\label{ssec:expressive}
In expressive speech, the voice production may differ significantly from the modal (or normal) phonation. As articulation may completely be changed, the vocal tract function can be subject to important modifications. In parallel, much of the dynamic variation in voice quality is brought about changes in phonation type and, hence, changes in the glottal source signal (\cite{Laver80}). As a consequence, the production of expressive speech is expected to be also reflected by relevant alterations in the glottal contribution.

In (\cite{Monzo}), authors investigate the use of speech-related and glottal features to discriminate between five expressive speech styles: neutral, sad, happy, sensual and agressive. The considered glottal parameters are the shimmer, jitter (see Section \ref{ssec:biomedical}) and the Glottal-to-Noise Excitation (GNE) ratio, and are observed to be of interest for this purpose. In (\cite{Sun_Emo}), the issue of differenciating emotions with a similar prosody is tackled by the consideration of glottal parameters. It is shown that the use of single glottal features reduced classification error for 24 emotion pairs in comparison to pitch or energy. In (\cite{Tahon}), the standard jitter and shimmer coefficients are complemented with the relaxation coefficient (Rd, estimated from the LF model) and the Functions of Phase-
Distortion (FPD) which are both related to the glottal production. Results conclude that glottal and usual voice quality features are of interest for the detection of the emotional valence. In (\cite{Eva}), authors tackle the problem of expressive speech style clustering with the goal of high-quality text-to-speech from audiobooks. For this, they make use of a set of standard glottal features derived from the LF model (Oq, Sq and Rq) which are further clustered with a Self-Organising Feature Map (SOFM).

As a particular expressive style, the glottal anaysis of the Lombard effect has also received attention in the literature. The Lombard reflex refers to the speech changes due to the immersion
of the speaker in a noisy environment. In such a context, the modification of the residual signal was investigated in (\cite{Bapineedu}) by considering features at the subsegmental level: the strength of excitation and a loudness measure reflecting the sharpness of the impulse-like excitation at GCIs. In (\cite{Drugman_Lombard}), the modifications of the glottal flow in Lombard speech were studied. For this, the glottal flow is estimated by a closed-phase analysis and parametrized by a set of time and spectral features. Significant and coherent changes of these parameters were observed as a function of the noise type and level. Also related to changes in speech intelligibility, the modifications of glottal and vocal tract information in hypo and hyperarticulation have been investigated in (\cite{Picart}). Increasing efforts of articulation were noticed to be associated with higher values of the pitch, the frequency of the glottal formant, the maximum voiced frequency and by a glottal flow containing a greater amount of high frequencies. 

Besides the aforementioned analysis and detection studies, several attempts have targeted the synthesis of expressive speech. The relevance of three speech components (spectral envelope, residual excitation and prosody) for synthesizing identifiable emotional speech has been estimated in (\cite{Barra}). Results proved the importance of transforming residual excitation for the identification of emotions that are not fully conveyed through prosodic means. In (\cite{Govind}), authors focused on the modification of the LP residual signal for emotion conversion. For this, the strength of excitation was modified by scaling the Hilbert envelope (HE) of the LP residual. The target emotion speech was finally synthesized using the prosody and strength modified excitation signal. The approach described in (\cite{Jaime}) aims at developing a HMM-based speech synthesis system with a controllable glottal source to manipulate the expressivity of the generated speech. The proposed approach builds on the GlottHMM parameters introduced in (\cite{Raitio}) and its viability is first analyzed by verifying that expressive nuances are captured by these features. The sensitivity of the method to speaker and recording conditions is also studied. Finally, remind that several techniques of parametric speech synthesis which have been shown to modify the produced voice quality have been presented at the end of Section \ref{ssec:synthesis}.

\section{Conclusion}
\label{sec:conclu}

The goal of this paper was to provide a comprehensive review of the advances made in glottal source processing. Starting from the fundamental and necessary tools of synchronization (pitch tracking and GCI detection), we then presented the existing methods for glottal flow estimation and parameterization. The last part of the paper finally described how the glottal-based information can be succesfully integrated within several voice technology applications: speech synthesis, speaker recognition, biomedical applications and expressive speech analysis.

The advantage of glottal source processing is that the excitation signal is expected to be highly complementary with features conventionnaly used in current voice technology systems, and which mainly characterize the vocal tract response. However this is done at the expense of an increase of the complexity of analysis techniques. Indeed, these latter techniques require the precise knwoledge of the fundamental frequency and/or of the GCI location, contrastingly to the conventional asynchronous MFCC or LPC-like feature extraction scheme. Another issue with glottal source estimation is that, since neither the vocal tract response nor the glottal contribution are observable, it is a typical blind separation problem. In other words, although some apparatus such as electroglottography (EGG) or high-speed imaging camera can be used to provide some information of the vocal folds behaviour, the glottal flow actually produced by speakers is unknown and no reference is available for validating the efficiency of glottal source estimation methods. Nevertheless, EGG recordings can be used as a ground truth for the development of synchronization tools: pitch tracking and GCI detection. Finally, the weakest point of current glottal source processing algorithms is related to their sensitivity and low robustness. Most of the developed approaches focus on the analysis in well controlled situations: synthetic signals, sustained vowels, studio recordings, etc. Although some progress has been made in this direction, further investigations are needed to allow a proper estimation of glottal information in realistic environments (noisy and/or reverberant) on connected speech. Analysis of high-pitched voices (e.g. for children), expressive speech and conversational data are also among the challenges to tackle over the next decades.


\section{Acknowledgements}
\label{sec:acknow}
Thomas Drugman is supported by FNRS.


\bibliographystyle{elsarticle-harv}
\bibliography{bibliography}







\end{document}